\documentclass[longauth]{aa}
\usepackage[varg]{txfonts}

\usepackage{graphicx}
\usepackage{comment}

\usepackage{natbib}

\def\apjl{The Astrophysical Journal, Letters}%
\def\apjs{ApJS}%
\newcommand{\pmi}{$\pm$}

\newcommand{\grados}{$^{\circ}$}
\newcommand{\fermi}{\emph{Fermi}-LAT}
\newcommand{\gbm}{\emph{Fermi}-GBM}
\newcommand{\agile}{\emph{AGILE}}
\newcommand{\maxi}{\emph{MAXI}}
\newcommand{\bat}{\emph{Swift}-BAT}
\newcommand{\integral}{\emph{INTEGRAL}}
\newcommand{\rxte}{\emph{RXTE}-ASM}
\newcommand{\xmm}{\emph{XMM-Newton}}

\newcommand{\cyg}{Cygnus\,X-1}
\newcommand{\cygt}{Cygnus\,X-3}
\newcommand{\pass}{{\sffamily{Pass 8}}}
\newcommand{\passseven}{{\sffamily{Pass 7 Reprocessed}}}

\begin{document}

\title{Detection of gamma rays of likely jet origin in \cyg}

%
\author{
R.~Zanin\inst{1} \and
A.~Fern{\'a}ndez-Barral\inst{2} \and
E.~de O\~na Wilhelmi\inst{3} \and
F.~Aharonian~\inst{1,4,5} \and
O.~Blanch\inst{2} \and
V.~Bosch-Ramon\inst{6} \and
D.~Galindo\inst{6} 
}
\institute{Max-Planck-Institut fur Kernphysik, P.O. Box 103980, D
  69029 Heidelberg, Germany
\email Roberta.Zanin@mpi-hd.mpg.de 
\and Institut de Fisica d'Altes Energies (IFAE), The Barcelona Institute of Science and Technology (BIST), Campus UAB, 08193 Bellaterra (Barcelona), Spain
\and Institute of Space Sciences (CSIC-IEEC) 08193 Barcelona Spain
\and Dublin Institute for Advanced Studies, 31 Fitzwilliam Place, Dublin
2, Ireland 
\and Universitat de Barcelona, ICC, IEEC-UB, E-08028 Barcelona, Spain
}


\abstract {} 
{Probe the high-energy ($>$60\,MeV) emission from the black hole X-ray
  binary system, \cyg, and investigate its origin.}
{We analysed 7.5\,yr of data by \fermi{} with the latest \pass{} software
version.}
{We report the detection of a signal at $\sim$8\,$\sigma$ statistical significance
  spatially coincident with \cyg{} and a luminosity above 60\,MeV of
  5.5$\times$10$^{33}$\,erg\,s$^{-1}$. The signal is correlated with the
  hard X-ray flux: the source is observed at high energies
  only during the hard X-ray spectral state, when the source is known
  to display persistent, relativistic radio emitting jets. The energy
  spectrum, extending up to $\sim$20\,GeV without any sign of spectral
  break, is well fitted by a power-law function with a photon index of
  2.3\pmi0.2. There is a hint of orbital flux variability, with
high-energy emission mostly coming around the superior conjunction.}
{We detected GeV emission from \cyg{} and probed that the emission is
  most likely associated with the relativistic jets. The evidence of flux
  orbital variability points to the anisotropic inverse Compton
  on stellar photons as the mechanism at work, thus constraining the
  emission region to a distance 10$^{11}-10^{13}$\,cm from the black hole. }

		
\maketitle

\section{Introduction}
\cyg{} is an X-ray binary (XRB), a system in which the compact object
accretes matter from the companion star. The former has been
identified as a black hole (BH) with (14.8\pmi1.0)\,M$_{\odot}$
\citep{Orosz:2011}. The latter is the early-type O9.7Iab supergiant
HDE\,226868 \citep{Walborn:1973}, with a mass of
(19.2\pmi1.9)\,M$_{\odot}$ \citep{Orosz:2011}. However, this value has
been questioned by \citet{Ziolkowski:2014}, who suggested a range of
25--35\,M$_{\odot}$. \cyg{} is the only high-mass XRB for which
the compact object has been identified to be a BH. 

Located at a distance of 1.86\,kpc \citep{Reid:2011,Xiang:2011}, it
is one of the brightest X-ray sources, thus considered an optimal
candidate for the study of the accretion and ejection processes onto a
BH system.
The spectrum of the BH X-ray binaries can be roughly described as the
sum of two components: a blackbody-like emission coming from the
geometrically thin, optically thick accretion disk, and a power-law
tail whose origin is still under debate.
The dominance of one or the other component defines the two main
spectral states the system can display: the soft state (SS) and the
hard state (HS).
The two main states are joined by short-lived (typically of a few days;
\citealt{Grinberg:2013}) intermediate states (IS), and
the complete sequence of states is well-represented in a hardness
intensity diagram (HID) by the q-shaped track trajectory
\citep{Fender:2004}. \\
The SS is dominated by the thermal emission peaking at $\sim$1\,keV
and a steep power-law at higher energies with photon index $\Gamma \sim$\,2--3.
In the HS instead, the blackbody component is much less luminous, with a
0.1\,keV temperature and most of the energy is emitted in a hard
tail component characterised by a $\sim$1.5 photon index and an
exponential cutoff at a few hundred keV. The canonical explanation for
this hard X-ray emission is inverse Compton scattering of disk photons
by hot ($kT_{e} \sim 100$\,keV) thermal electrons in the inner region
of the accretion flow, usually referred to as ``corona'' 
\citep{Shapiro:1976,Sunyaev:1979}. 
However, \citealt{Aharonian:1985} proposed that this
emission has a non-thermal origin related to the development of
electromagnetic cascades initiated by particles accelerated to
relativistic energies in regions close to the BH, i.e. in the
accretion disk. In this scenario the authors showed that the resulting
photon spectrum has a spectral break at most at $\sim$1\,MeV. 
In addition, the HS generally displays relatively persistent relativistic 
jets emitting synchrotron radiation at GHz radio
frequencies, whereas in the SS, where the disk comes up to the BH
last stable circular orbit, such an emission is strongly quenched. 
A two-cluster non-linear correlation between the radio and the X-ray
fluxes, with slopes of $\sim0.7$ and $\sim$1, respectively, 
suggests that there is a close coupling between the X-ray and the
radio emitters \citep{Gallo:2003, Gallo:2012}.  The existence of
such a correlation was used to prove a possible synchrotron origin of the
X-ray power-law tail \citep{Markoff:2003}. 

As a persistent source, \cyg{} retains always a strong power-law
spectral component; even in its SS, i.e. its spectrum is never fully
disk-dominated. 
Whereas in the HS it shows a mildly relativistic ($v \sim0.6\,c$) radio jet 
\citep{Stirling:2001,Gallo:2003} which carries a significant fraction of
the system X-ray luminosity \citep{Gallo:2005:nature}, in the SS
there is evidence for a factor 3--5 weaker unresolved compact jet
\citep{Rushton:2012}. The constant mean level of the radio emission is
of $\sim$10--15 mJy, with a flat spectrum and no evidence for a cutoff 
\citep{Fender:2000} up to IR frequencies, where the emission is totally
dominated by the supergiant, making impossible the measurement of the
spectral break. 
Another peculiarity of the \cyg{} HS is that above the 
hard X-ray tail which cuts off at $\sim$100\,keV \citep{Wilms:2006},
an additional harder (with a 1.6 photon index) non-thermal component
emerges extending up to a
few MeV \citep{Cadolle:2006,Rodriguez:2015}.
This soft gamma-ray radiation was recently shown to be polarised with a
polarisation fraction increasing with energies and an average value of
(76\pmi15)\% at a position angle of (42\pmi3)\grados{} for energies
above 230\,keV \citep{Laurent:2011,Jourdain:2012}.  The most plausible
explanation for this is that the jet
synchrotron emission extends itself up to MeV energies
\citep{Jourdain:2012,Zdziarski2012}, requiring the
existence of a parent population of ultra-relativistic electrons. In addition, such a 
high-level of polarisation would imply that the phenomenon is
persistent on time-scales of weeks to months. However, the origin of
this emission is still controversial \citep{Zdziarski:2004} and it was 
also suggested to originate in the corona \citep{Romero:2014}.
Therefore, the only proof that there is non-thermal gamma-ray jet
emission would undoubtedly be the detection of GeV emission.

\cyg{} spends most of its time in the HS, although the fraction of time
observed in the SS is not constant with time. The latter increased
from 10\% between 1996 and 2000, to 34\% between 2000 and mid 2006
\citep{Wilms:2006}, most probably due to an overall increase of the
stellar radius. 
In this work we adopt the state definition described in
\citet{Grinberg:2013} who used the data of the available all-sky monitors:
\rxte, \maxi, \bat, and \gbm. They showed that publicly available
\bat{} (15--50\,keV) data can be used to distinguish SS from the
HS+IS: the source is in the SS when the \bat{} daily count rate is
smaller than 0.09\,cts\,cm$^{-2}$\,$\mathrm{s}^{-1}$. However, without
soft coverage it is not possible to distinguish between the HS and the
IS. 

\begin{figure}[!h]
\centering
\includegraphics[width=0.4\textwidth]{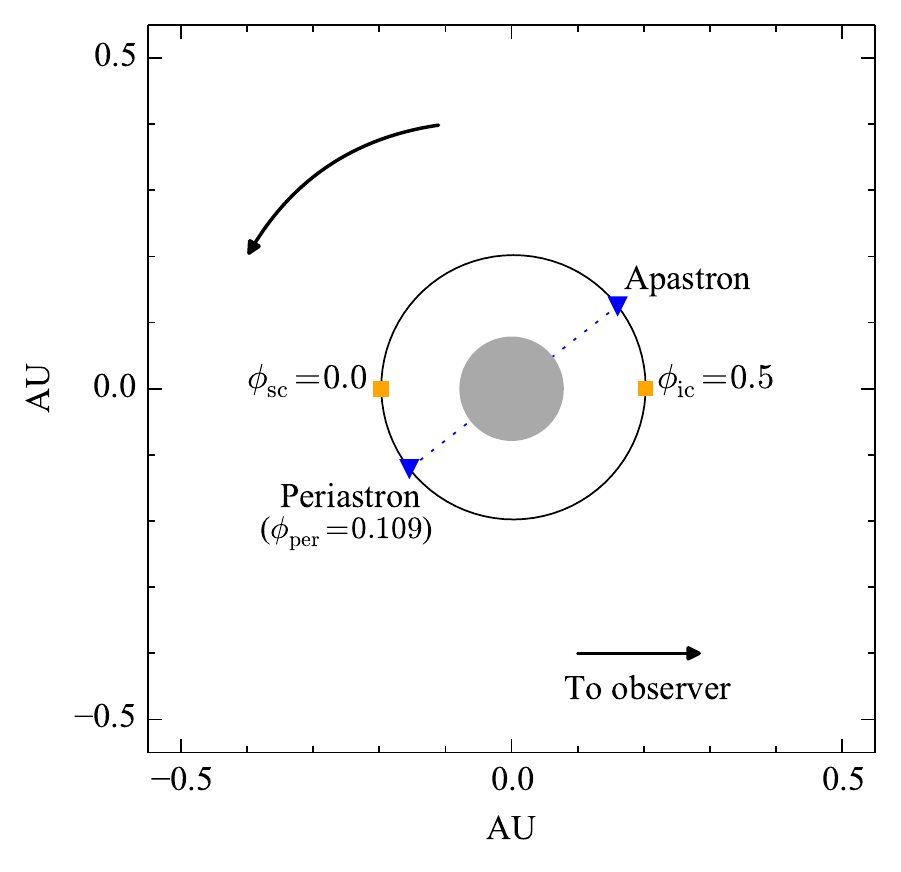}
\caption{Schematic diagram of \cyg. The orbital phase $\phi$=0
  corresponds to the superior conjunction of the compact object. The
  orbit of the BH is indicated by the ellipse with an eccentricity of
  0.018 \citep{Orosz:2011} . Neither the inclination of the orbit with the
  line of sight or the longitude of the ascending node were
  considered. The stellar
  radius, assumed to be 16.4R$_{\odot}$, as well as 
  the periastron phase and argument were taken from
  \citet{Orosz:2011}. AU stands for astronomical unit.
\label{fig:orbit}}
\end{figure}

\cyg{} BH is in a 5.6\,d orbit. In this paper we adopted the most
updated ephemerides in \citet{Gies:2008}, with a phase 0 corresponding
to superior conjunction T$_0$=52872.788 HJD, i.e. when the
companion is between the observer and the BH (see the schematic
diagram of the binary in Figure\,\ref{fig:orbit}). 
The orbital period is observed at all wavelengths:
optical, infrared \citep{Gies:1982}, X-ray \citep{Brocksopp:1999}, and
also at radio frequencies \citep{Pooley:1999}, suggesting that such a
modulation could be the result of absorption by the stellar wind
\citep{Brocksopp:2002}. As confirmation, \citet{Grinberg:2015} show
that the absorption column density in the HS is strongly modulated 
with a maximum around superior conjunction. 
The existence of the radio modulation supports the idea that radio
emission comes from a continuous jet versus discrete ejections.  
Another type of periodical behaviour is observed in \cyg{} both at X-ray
and radio frequencies, a superorbital modulation of $\sim$140\,d
\citep{Brocksopp:1999,Pooley:1999}, although such a value is
rather unstable and it has been recently showed to be doubled
\citep{Lachowicz:2006,Rico:2008,2011MNRAS.412.1985Z}. 
The superorbital modulation is
possibly related to the precession of the disk-jet system
\citep{Brocksopp:1999}, or alternatively to a variable mass accretion
rate \citep{Brocksopp:2001}, and its period possibly varies when an
X-ray spectral state change occurs \citep{Rico:2008}. 

HE emission from BH XRB is theoretically predicted invoking either
leptonic or hadronic processes (e.g. \citealt{Bosch-Ramon:2009} for a
review) and generally tied to the existence of the radio jets, where
particles can be accelerated up to relativistic energies. So far, the
only microquasar (i.e. XRB displaying relativistic jets) firmly
detected at high energies ($>100$\,MeV) is \cygt{}
\citep{Tavani:2009,Abdo:2009} and its gamma-ray emission is related to
the formation/existence of the radio jets \citep{Piano:2012,Corbel:2012}. 
However \citet{Malyshev:2013} showed there is a
4$\sigma$-level evidence of gamma-ray signal for \cyg, above 100\,MeV,
in 3.8\,yr of \fermi{} data, only when the source is in the HS. 
In addition to this steady emission there are claims of isolated
1-2-day long flaring events reported by \agile{} above 100\,MeV
\citep{Sabatini:2010,Sabatini:2013} and of a flare of $\lesssim 1$\,d
duration reported by the MAGIC collaboration above
100\,GeV \citep{Albert:2007}. In particular, \agile{} detected three
episodes of significant transient emission while it was
in its pointing operational mode. These 1--2\,d long events occurred
on 2009, October 16 (0.38--0.56 orbital phase) with an integral flux of
(2.32\pmi0.66) $\times 10^{-6}$\,ph\,cm$^{-2}$\,s$^{-1}$ 
between 0.1 and 3\,GeV \citep{Sabatini:2010}; on 2010, March 24
\citep{Bulgarelli:2010} with a flux above 2.50 $\times
10^{-6}$\,ph\,cm$^{-2}$\,s$^{-1}$ above 100\,MeV; and on 2010, June 30 with a
(1.45\pmi0.78) $\times 10^{-6}$\,ph\,cm$^{-2}$\,s$^{-1}$ average flux in the
same energy range \citep{Sabatini:2013}. Whereas the first two episodes
happened when the source was in the HS, the last one occurred during a
hard-to-soft state transition, but coincident with
the source entering in the SS and a couple of days before of an
anomalous radio
flare \citep{Negoro:2010,Rushton:2010,Wilson:2010}. 
An independent analysis of 3.6\,yr of \fermi{} data confirmed evidence
of flaring activity on 1--2\,d timescales contemporaneous, but not
coincident, with \agile{} at 3--4$\sigma$ level \citep{Bodaghee:2013}.
Such episodes show integral fluxes typically lower than the ones
reported by \agile{}, but still compatible within the large statistical uncertainties. 
The reported evidence (at 4$\sigma$ post-trial) of very-high-energy
(VHE, $>$100\,GeV) emission was detected by MAGIC on 2006, September
24 for 80 minutes (corresponding to an orbital phase of 0.9) when the
source was in the HS \citep{Albert:2007}, but it occurred exactly one
day before of a hard X-ray flare observed by \integral{}
\citep{Malzac:2008}. Further long observational campaigns
($\sim$100\,hr) were carried out by MAGIC meant to catch additional short
flaring episodes similar to the September-2006 one, but with no
success \citep{Alba:2015}. VERITAS did not report any VHE signal
from \cyg{} too \citep{Guenette:2009}. \\
In this work we search for both steady and variable emission from
\cyg{} at high energies, above 60\,MeV, by using 7.5\,yr of data by
\fermi.

\section{Observations and analysis}

\fermi{} is an electron–positron pair production telescope, featuring
solid state silicon trackers and cesium iodide calorimeters, designed
to be sensitive to photons from $\sim$20\,MeV up to $>300$\,GeV
\citep{Atwood:2009}.

We used 7.5\,yr of \pass{} \fermi{} data from August 4, 2008 (MJD\,54682)
to February 2, 2016 (MJD\,57420). The recently released \pass{} data
benefits for a wider energy range (from 60\,MeV to 500\,GeV), better
energy resolution, improved point spread function (PSF), and
significantly increased effective area. In addition, more accurate
Monte Carlo simulations of the detector led to a reduction of the
systematic uncertainty in the LAT instrument response functions
(IRF). 
The data were reduced and analysed using
\emph{Fermipy}\footnote{http://fermipy.readthedocs.org/en/latest/}, a
set of python tools which automatize the \pass{} analysis with the FERMI
SCIENCE TOOLS v10r0p5 package\footnote{See the Fermi
  Space Science Center (FSSC) Web site for details of the Science Tools:
  http://fermi.gsfc.nasa.gov/ssc/data/analysis/software}. Standard
FERMI SCIENCE TOOLS were also used to cross-check the recently
released python package.
We selected LAT photons from the ``P8R2\_SOURCE'' class in the 
 widest possible energy range between 60\,MeV and 500\,GeV and within
 a 30\grados{} acceptance cone centered on the position of \cyg{}
(RA$_{\mathrm{J2000}}$=19h58m21.676s, Dec=+35\grados{}12m5.78s,
\citealt{vanLeeuwen:2007}). We chose 60\,MeV as the lowest energy threshold
because that is the minimum energy of the public Galactic diffuse
model.  The ``SOURCE'' event class was chosen to
maximize the effective area, knowing that \cyg{} is expected to be
pointlike. Thus, we used the corresponding ``P8R2\_SOURCE\_v16''
IRF.
Within any photon class, \pass{} subdivides the events into
quartiles according to the quality of either the direction (PSF
event-type) or the energy reconstruction (EDISP event-type). In this
work, we analysed the four PSF event-types separately and later we
combined them by means of a joint likelihood fit.
In order to minimize
the contamination by the albedo gamma rays from the Earth, we excluded those photons having
reconstructed directions with angles with respect to the local zenith
larger than  90\grados, 85\grados, 75\grados, and 70\grados{} for the four PSF quartiles, respectively:
the tighter the better the PSF is. Since our analysis is not very sensitive to
the contamination from the Earth Limb, and, in addition, our data
sample includes data taken during the period of the Galactic-Center
biased pointing strategy, we did not apply any cut on the rocking angle. 
The PSF event-type analysis has the best achievable angular resolution
among the possible \fermi{} analysis, e.g. at 1\,GeV it is
$\sim$0.5\grados, being a value between 0.8\grados{} and 0.3\grados{}
which are the angular resolutions of the analysis that uses the
four PSF event-type in one single likelihood and of the one of the
quartile with the best PSF (PSF3),
respectively\footnote{https://www.slac.stanford.edu/exp/glast/groups/canda/lat\_Performance.htm}.
We cross-checked those results repeating the analysis with the
standard conversion-type (FRONT+BACK) selection of events, and a
conservative maximum zenith of 90\grados. 
The results obtained with the two selections are compatible within the
expected improvements of the first approach.

We created 14\grados{} $\times$ 14\grados{} regions of interest (ROI)
in Galactic coordinates.
To model the diffuse background, we used the templates for the
Galactic diffuse emission (gll\_iem\_v06.fits) and an isotropic
component (iso\_P8R2\_SOURCE\_V6\_PSFx\_v06.txt with
x=0,1,2,3) including the extragalactic diffuse
emission and the residual background from cosmic
rays\footnote{http://fermi.gsfc.nasa.gov/ssc/data/access/lat/BackgroundModels.html}.
We built a model of point-like gamma-ray background sources within
22\grados{} starting with the third LAT source catalogue (3FGL, 
\citealt{Fermi:2015:3FGL}), which is based on 4\,yr of \passseven{}
data. We let all the spectral parameters of all the sources, except
those located more than 7\grados{} away from the center, free to
vary in a maximum likelihood fit (using \emph{gtlike}). We also let free
the flux normalisation of the Galactic
diffuse and isotropic components and of the extremely bright 3FGL
sources (significance $>$100) lying between 7\grados{} and 14\grados{}
from the center.

Source detection significance is determined, fixing the source
  position to its nominal value given by \citet{vanLeeuwen:2007},
  using the Test Statistic value, TS =$-2\ln$(L0/L1) which compares
  the likelihood ratio of models including, e.g., an additional
  source, with the null hypothesis of background only \citep{Mattox:1996}. 
The TS maps were computed for a power-law test source with a photon
index of 2.5 and obtained with all the background sources fixed.
The TS maps presented in this work were obtained above 1\,GeV where the
angular resolution is $\sim$0.5\grados{}.

For the spectral analysis we splitted the 0.06-500\,GeV energy range
into 7 logarithmically spaced bins. 
The spectral energy distribution (SED) was computed by fitting the
source normalisation factor in each energy bin independently while
keeping its photon index fixed to the value found in the
overall, full energy range, fit. The spectral parameters of the
background sources were fixed to those previously found in the overall
fit.
For each spectral point we required a TS of at least 4, when this
condition was not fulfilled, upper limits (UL) at 95\% confidence
level (CL) were computed. 

The source localisation was performed above 1\,GeV with a two-step algorithm: first
it looks for the maximum peak in a reduced TS map of
4\grados\,$\times$\,4\grados{} centered around the considered source,
and then it redefines the source position by performing a full
likelihood fit in the vicinity of the peak found in the first
step\footnote{http://fermipy.readthedocs.org/en/latest/advanced/localization.html}, 
with the normalisation parameters of bright (TS$>$100) background
sources let free to vary. 

The light curve, i.e. integral flux as function of the
observation time, is the only result which was not produced with the
\emph{Fermipy} software package, but with the standard FERMI SCIENCE
TOOLS and with the standard conversion-type selection of events and a
maximum zenith angle cut of 90\grados. 
The energy threshold for the light curve is 100\,MeV in order to have a
direct comparison with the previously published results
\citep{Sabatini:2010,Sabatini:2013,Bodaghee:2013}. We computed a
maximum likelihood fit for each temporal bin of 1 \,d, and then we
estimated either its 0.1--20\,GeV integral flux or 95\% CL UL
depending on the strength of the signal, with a threshold of TS=9. 

\section{Results}
The TS map above 1\,GeV obtained by using the background model
including all the 3FGL sources is not flat. Besides a clear excess in the center,
coincident with \cyg{}, the TS
map shows 7 excess spots with a TS larger than 25 in the
full energy range, between 60\,MeV and 500\,GeV (see Figure
\ref{fig:TSbg}). We modelled these excess spots as power-law
point-like sources, in particular:
\begin{figure}[!h]
\centering
\includegraphics[width=0.4\textwidth]{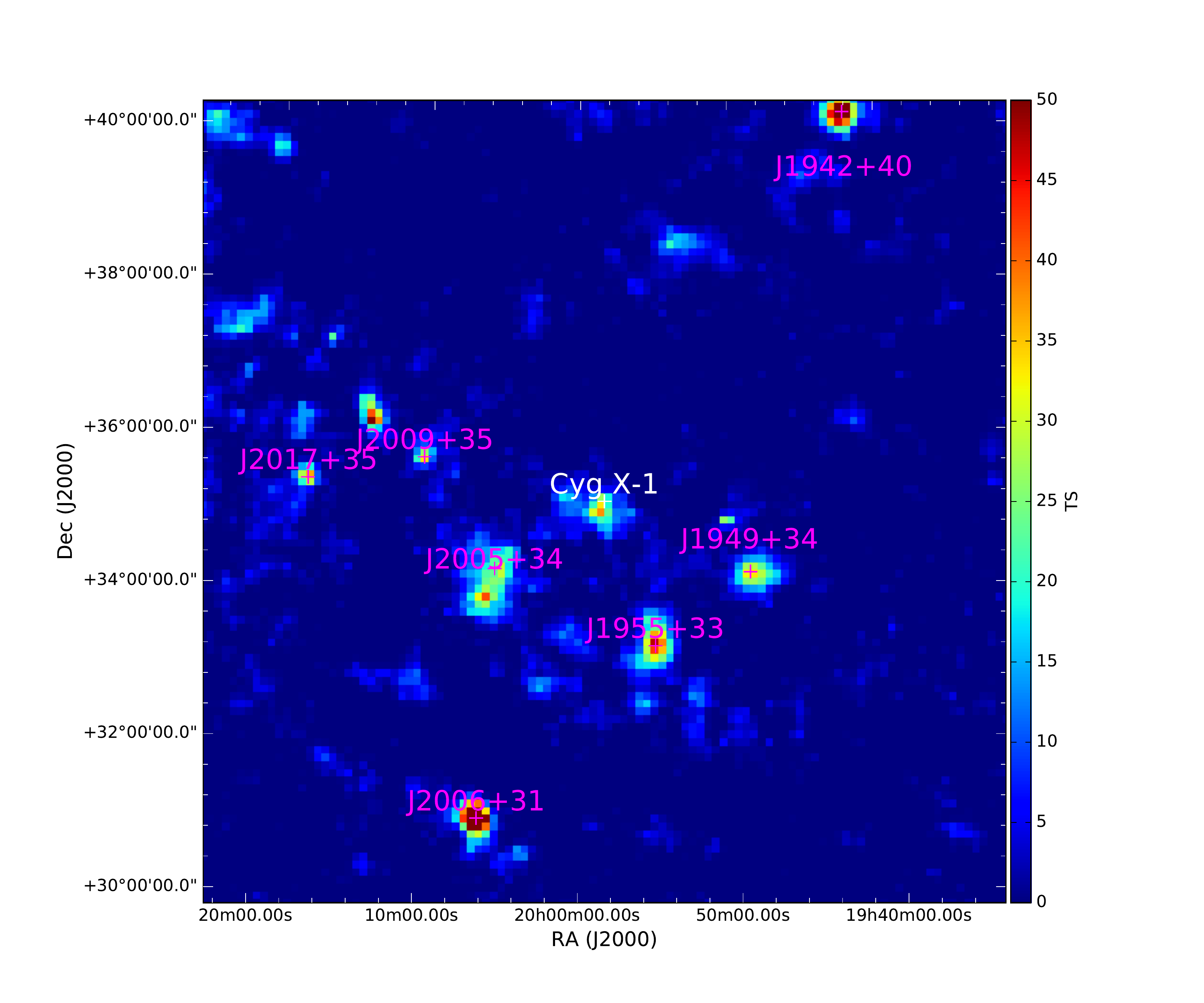}
\caption{5\grados\,$\times$\,5\grados{} TS map centered at the
  nominal position of \cyg{} above 1\,GeV obtained when including only
  the 3FGL sources in the background model.
\label{fig:TSbg}}
\end{figure}
\begin{itemize}
\item
J1942+40: the LAT excess located at
RA$_{\mathrm{J2000}}$=19h:42m:7s and
Dec=+40\grados:14m:7s most probably comes from the
direction of the open cluster NGC\,6819 where several X-ray sources
were detected by the \xmm{} observatory \citep{Gosnell:2012}. In the
full energy range the source has a TS of 55. 
\item
J1949+34: a LAT excess with a TS of 35 in the full energy range
which is located at RA$_{\mathrm{J2000}}$=19h:49m:7s and
Dec=+34\grados:15m:44s. 
\item
J1955+33: an excess located at  RA$_{\mathrm{J2000}}$=19h:55m:10s,
Dec=+33\grados:18m:34.8s, and with a TS of 90 in the full energy range.
\item
J2005+34: a LAT excess centered in RA$_{\mathrm{J2000}}$=20h:05m:19.7s,
Dec=+34\grados:18m:23.7s with a TS of 49 above 60\,MeV.
\item
J2006+31: a clearly identified (TS=115) new LAT source,
outside of the Galactic Plane, it is centered in
RA$_{\mathrm{J2000}}$=20h:06m:12.8s, Dec=+31\grados:02m:38.3: is
spatially coincident with the 164\,ms period radio pulsar PSR\,J2006+3102
\citep{Nice:2013}. We found that the LogParabola function provides a
better (at more than 3\,$\sigma$ level) fit of this source
spectrum. Just for this specific case we included a non-power-law
spectrum in our background model. 
\item 
J2009+35: a LAT excess with a TS of 48 (above 60\,MeV) located at
RA$_{\mathrm{J2000}}$=20h:09m:57.8 and Dec=+35\grados:44m:48.6s. 
\item 
J2017+35: a LAT excess with a TS of 65, above 60\,MeV, located at
RA$_{\mathrm{J2000}}$=20h:17m:25s and Dec=35\grados:26m:5s.
\end{itemize}
The search for the origin of these excesses
goes beyond the goal of this paper.  Their location was estimated
above 1\,GeV and has a statistical uncertainty of $\sim$0.2\grados.
In addition, we found that the centroid of the LAT source associated
with the SNR\,G73.9+0.9 is offset by 0.24\grados{} with respect to the
position given in the 3FGL catalogue (3FGL\,J2014.4+3606). The new centroid is
located at RA$_{\mathrm{J2000}}$=20h:13m:33.8s and
Dec=36\grados:11m:54.0s. The LogParabola
spectral model suggested by the new \pass{} analysis in
\citet{Zdziarski:2016} is not significantly favoured with respect to a
power-law function, which was used in this work. 

\begin{figure*}[!h]
\centering
\includegraphics[width=0.8\textwidth]{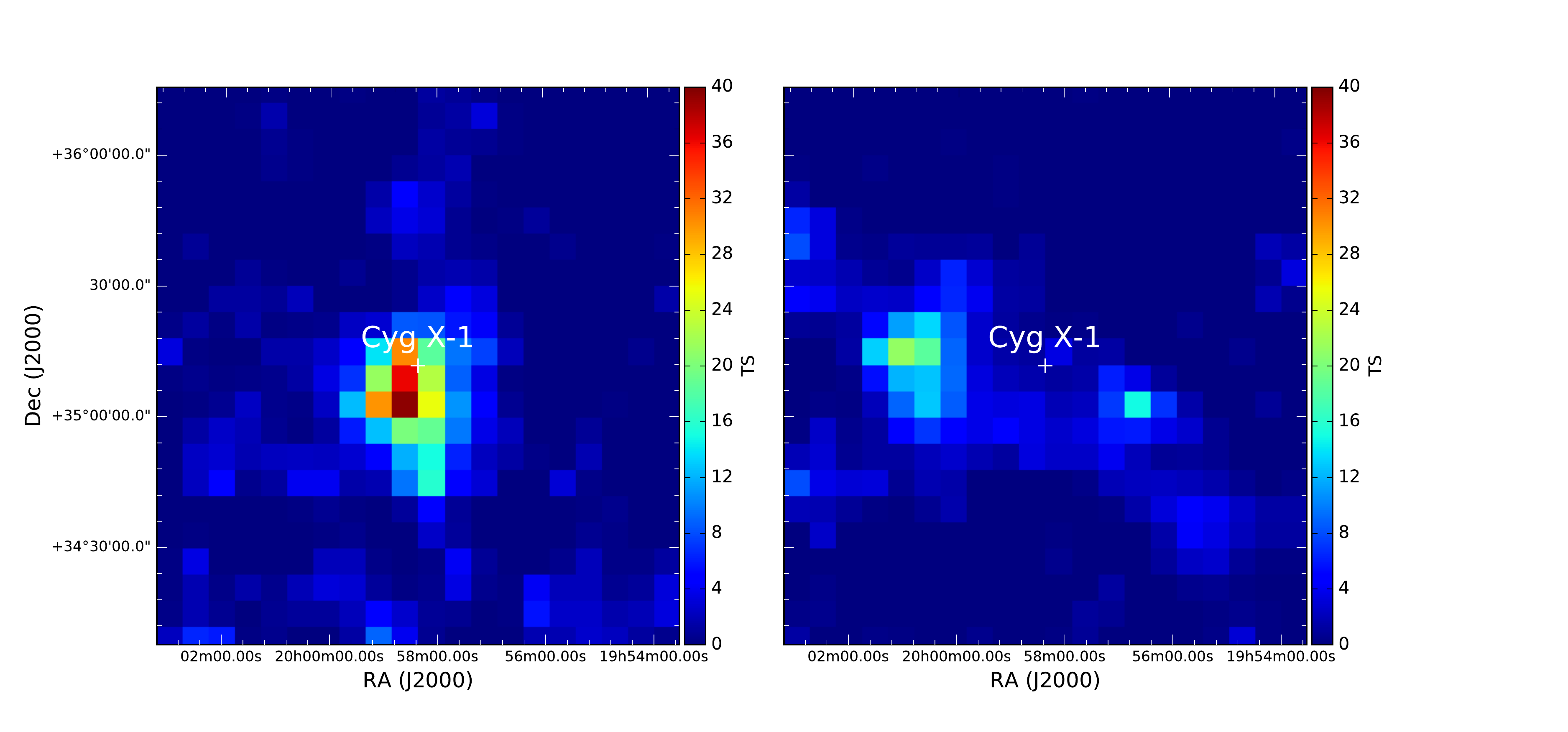}
\caption{2\grados\,$\times$\,2\grados{} TS maps centered at the
  nominal position of \cyg{} above 1\,GeV. \emph{On the left:} the
  \fermi{} data subsample corresponding to \cyg{} being in the HS;
  \emph{On the right:} the data subsample corresponding to the SS of
  the source. The white cross indicates the nominal position of
  \cyg. 
\label{fig:TSmapSSHS}}
\end{figure*}

\begin{figure*}[!h]
\centering
\includegraphics[width=0.8\textwidth]{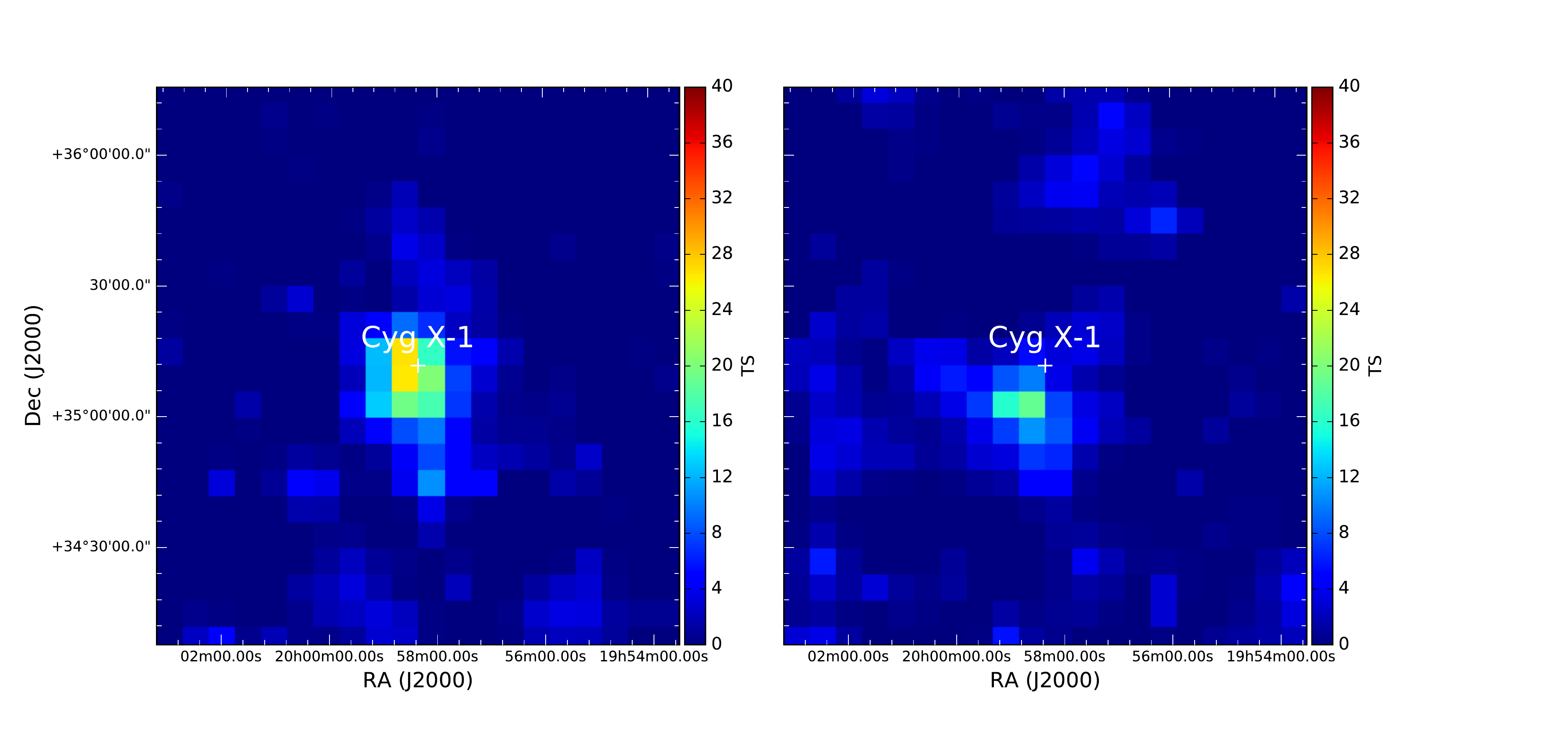}
\caption{Phase-dependent TS map of a 2\grados{} $\times$ 2\grados{}
  region centered on \cyg{} (above 1\,GeV) when the source is in the
  HS  \emph{on the left} around the superior conjunction and \emph{on
    the right} around the inferior conjunction . 
\label{fig:TSmapICSC}}
\end{figure*}

\begin{figure*}[!h]
\centering
\includegraphics[width=0.8\textwidth]{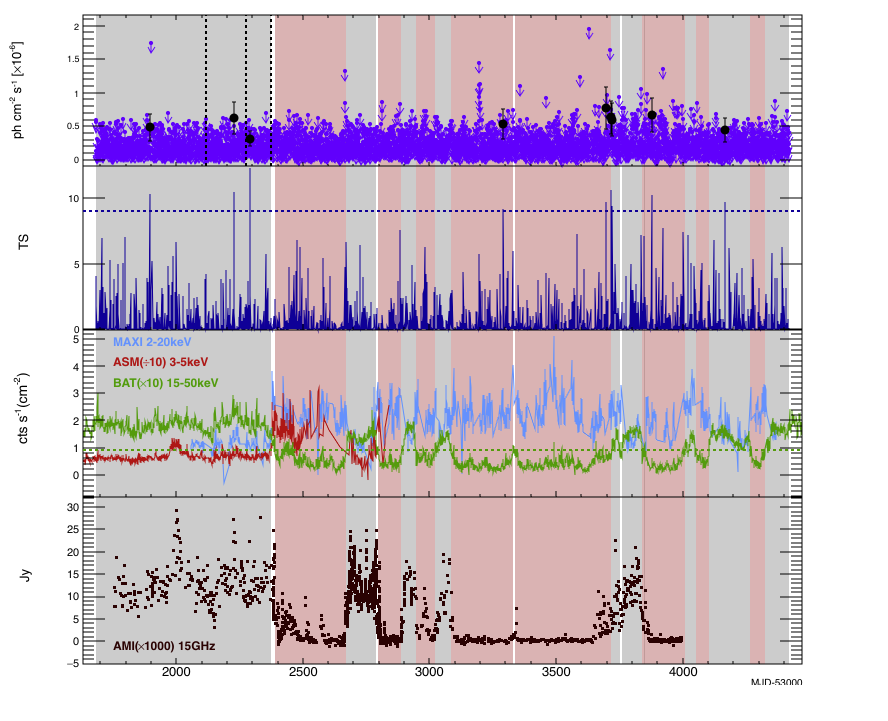}
\caption{Light curve of \cyg{} obtained by the analysis of the
  \fermi{} data between August 4, 2008 and February 2, 2016 in 1\,d
  bins. From top to bottom: the integral fluxes in the 0.1--20\,GeV
  energy range, the TS corresponding to the above integral fluxes, the
  daily light curves in soft and hard X-rays from \maxi
  (cts\,s$^{-1}$\,cm$^{-2}$ in 2--20\,keV), \rxte{} (cts\,s$^{-1}$ in
  3--5\,keV divided by 10) and \bat{} ($\times$10
  cts\,s$^{-1}$\,cm$^{-2}$ in 15--50\,keV), and the daily integral
  fluxes at 15 GHz by AMI. In the 0.1--20\,GeV range, days with TS$>$9 
  are shown in the top panel as filled black circles, whereas the
  purple arrows represent the 95\% CL UL for the bins with
  TS$<$9. Dashed black vertical lines show the previously reported
  gamma-ray flares by \agile. The horizontal green
  dot-dashed line in X-ray panel indicates the threshold level of the
  \bat{} count rate used in this work to separate the SS from the HS(+IS):
  0.09$\times$10\,cts\,s$^{-1}$\,cm$^{-2}$. The shadowed gray bands
  identify the HS(+IS), whereas the red ones the SS. 
\label{fig:LC}}
\end{figure*}

Once the new background sources are included in our model, a point-like
source at the position of \cyg{} is popping up at TS=53
in the full energy range, between 60\,MeV and
500\,GeV. Above 1\,GeV the source is still detected at TS of 31.
Among the new background sources included in our model the ones
at more than 3\grados{} away from the \cyg{} position do not have any
effect on the source significance estimation, whereas the new
  background excesses lying within a 3\grados{} radius from the
  nominal \cyg{} position decrease the TS of the signal from 65 to 53,
  once included in the background model. 
Applying the localisation algorithm in the \emph{Fermipy} package, we
fitted the position of the gamma-ray excess above 1\,GeV to 
 RA$_{\mathrm{J2000}}$=19h:58m:56.8s and Dec=+35\grados:11m:4.4s,
 0.05\grados{} offset from the nominal position, but still compatible
 with \cyg{} within the statistical uncertainties of 0.2\grados. 
Besides the spatial coincidence, we found a strong correlation of the
gamma-ray excess with the X-ray spectral states. We divided our sample
in HS(+IS) and SS by using the public \bat{} data (daily \bat{} count rate lower/larger than
0.09\,cts\,cm$^{-2}$\,$\mathrm{s}^{-1}$, \citealt{Grinberg:2013}),
which has a complete temporal overlap with \fermi.
We did not account for any soft X-ray information in our selection
criteria, aiming to clearly identify all the IS intervals, because,
given the short duration of the IS, their eventual
inclusion in the HS does not alter our result. The bottom panel in
Figure \ref{fig:LC} shows the X-ray fluxes
as a function of the time interval considered in this work (August
2008 - February 2016), and the two X-ray spectral states are emphasised
by the different-colored-shadowed bands. In particular, the HS and SS
intervals in MJD are listed in Table\,\ref{tab:mjd}.
We detected \cyg{} when it was in the HS with a TS of 49 above 60\,MeV
and an energy flux integrated over the entire energy analysis range of 
(7.7\pmi1.3)$\times10^{-6}$\,MeV\,cm$^{-2}$\,s$^{-1}$. On
the other hand, there is no significant LAT excess in coincidence with \cyg{}
when it is in the SS (TS=7) and the UL on its energy flux above
60\,MeV at 95\% CL is of 5.4$\times10^{-6}$\,MeV\,cm$^{-2}$\,s$^{-1}$.
The two corresponding TS maps for energies larger than 1\,GeV are shown in
Figure \ref{fig:TSmapSSHS}. 
Given the comparable exposure time for the two considered LAT
subsamples of 3.6\,yr and 3.7\,yr, for the HS and SS, respectively, we
do not need to normalize with respect to it and we can
confirm that \cyg{} is detected at high significance only when
  the source is in the HS,  as previously claimed by
\citet{Malyshev:2013}, and recently confirmed by
\citet{Zdziarski:2016:cyg}. Some high-energy emission in the SS, although
significantly fainter than in the HS, cannot be excluded.
Spectral and timing results were then computed
only for the HS subsample. 
We checked for a possible dependence on the orbital period of the flux. Given the
low significance of the signal we divided the HS data sample in 
only two bins, one centered on the superior conjunction 
($\phi>0.75\,||\,\phi<0.25$), and one on the inferior conjunction
($0.25<\phi<0.75$). \cyg{} is clearly detected in the full energy
range with a TS of 31 only around the
superior conjunction. In the second bin, around the inferior
conjunction, no significant signal from the \cyg{} position is seen
(TS=10). Figure \ref{fig:TSmapICSC} shows the corresponding two TS
maps above 1\,GeV. The energy flux above 60\,MeV is
(7.6\pmi1.7)$\times10^{-6}$\,MeV\,cm$^{-2}$\,s$^{-1}$ for the phase
interval around superior conjunction. The low statistics
does not allow us to make
any strong conclusion on a possible flux dependence on the orbital
position. Nevertheless, the low significance on the inferior
conjunction phase bin, for the same exposure time, can be considered
as a hint of the orbital modulation of the flux. 

\begin{table}	
\centering
\small
\begin{tabular} { c  c }
\hline
\hline
HS & SS \\
\hline
54682--55375 & 55391--55672 \\
55672--55790 & 55797--55889 \\
55889--55945 & 55945--56020 \\
56020--56086 & 56086--56330 \\
56718--56753 & 56338--56718 \\
56759--56839 & 56839--57009 \\
56848--56852 & 57053--57103 \\
57009--57053 & 57265--57325 \\
57103--57265 & \\
57325--57420 & \\
\hline
\end{tabular}
 \caption{\label{tab:mjd}
          Days (MJD) intervals for the HS and SS. }
\end{table}

\begin{figure}[!h]
\centering
\includegraphics[width=0.45\textwidth]{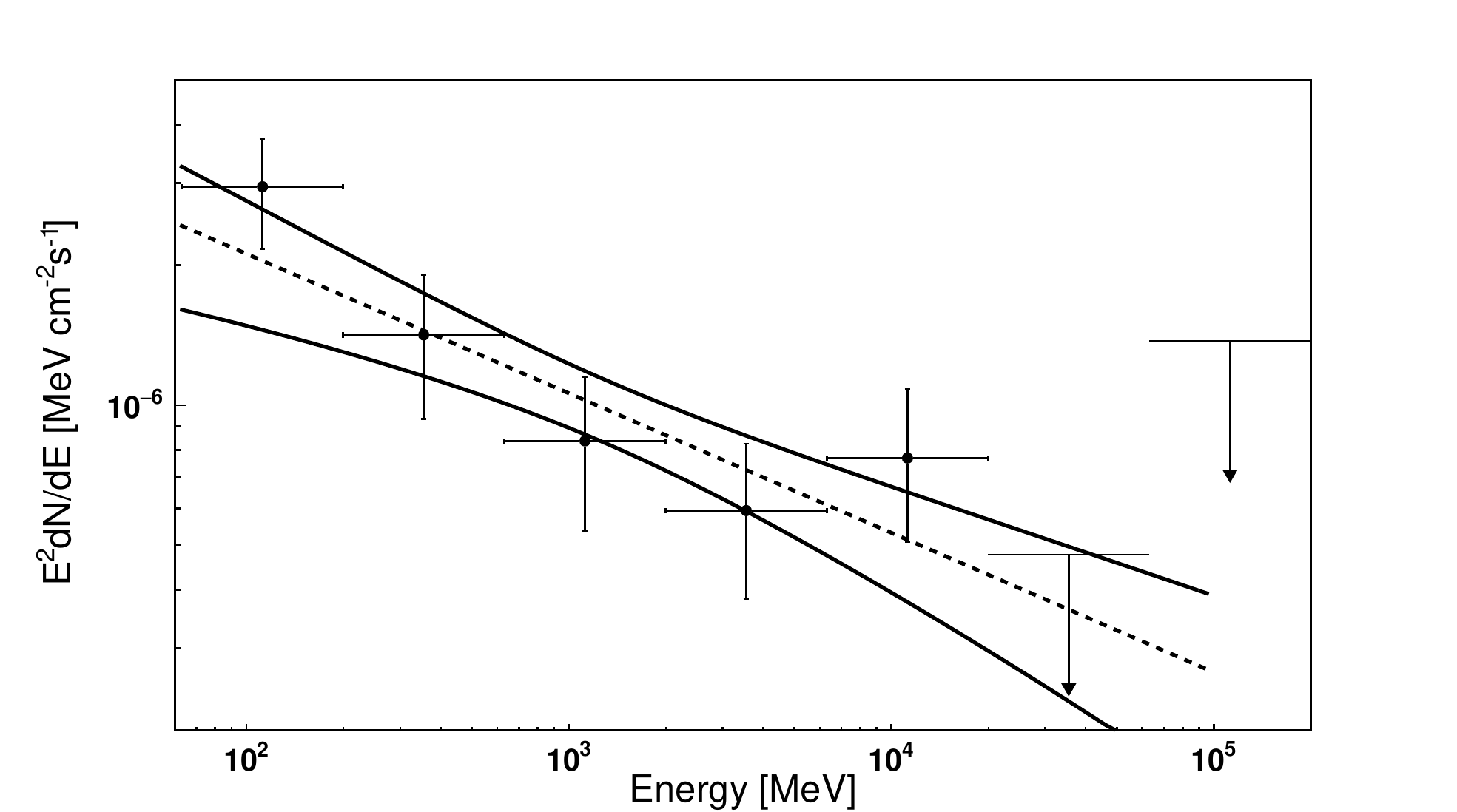}
\caption{SED of \cyg{} in the HS, extending from 60\,MeV to
  20\,GeV. It is best fitted by a power-law function with a photon
  index $\Gamma$=(2.3\pmi0.1).  
\label{fig:SED}}
\end{figure}

The energy spectrum of \cyg{} is well fitted by a power-law function
with a photon index $\Gamma$=(2.3\pmi0.1) and a normalisation factor of
$N_0$=(5.8\pmi0.9) $\times 10^{-13}$\,MeV$^{-1}$\,cm$^{-2}$\,s$^{-1}$
at the decorrelation energy of  1.3\,GeV, and it extends from 60\,MeV
to $\sim$20\,GeV. The
obtained SED is illustrated in Figure \ref{fig:SED}. We also tried to
fit the LAT data with a
broken power law, but the obtained improvement is not statistically
significant ($\Delta$TS$<$2). The photon indices of the inferior and
superior conjunction energy spectrum are compatible within 1\,$\sigma$
with the overall HS ones. The flux normalisation at 1.3\,GeV for the superior
and inferior conjunction, computed both with 2.3 photon index, is
of $N_0$=(5.7\pmi1.3) $\times
10^{-13}$\,MeV$^{-1}$\,cm$^{-2}$\,s$^{-1}$, and $N_0$=(3.7\pmi1.3)
$\times 10^{-13}$\,MeV$^{-1}$\,cm$^{-2}$\,s$^{-1}$, for the superior
and inferior conjunction, respectively.

We performed a timing analysis looking for the orbital period of 5.6
d, or eventually the super-orbital period. 
We applied the Lomb-Scargle test of uniformity \citep{Lomb:1976},  to
our daily light curve for the HS data sample, where significant
signal is detected, using the most probable value for the integral
flux regardless the TS values. Between the maximum and minimum period sampled 
(2 and 1000\,d), no periodicity is observed being the most significant
one compatible at 3\% with the null hypothesis of no periodicity.

We also looked for short flux variability on daily timescales as
claimed by the AGILE collaboration. Figure \ref{fig:LC} shows the light curve of
the 7.5\,yr of data with a one-day binning, in the 0.1--20\,GeV energy
range, where 20\,GeV is the limit to which \cyg{} energy spectrum
extends. No hint of variability at daily timescale is observed and the
distribution of the daily TS is consistent with the expected $\chi^{2}$
distribution. For completeness, Table \ref{tab:LC} lists the 9 days
showing signal with TS$>$9. They
are not clustered around previously reported daily variabilities,
being none of them coincident with the AGILE claims
\citep{Sabatini:2010,Sabatini:2013,Bulgarelli:2010} 
and just one (MJD\,55292) with a 3$\sigma$ event detected by
\citealt{Bodaghee:2013}.   
Nevertheless, possible differences on the exposure times and the
  effective area degradation due to large off-axis angles of the
  \cyg{} daily observations by the two gamma-ray detectors AGILE and
  \fermi{} could explain this apparent contradiction, as it shown for
  the case of AGL\,J2241+4454 \citep{munar:2016}.

\begin{table*}[ht]\label{tab:LC}\footnotesize
	\centering
	\begin{tabular}{ c  c c  c  c }
		\hline
		\hline
		 \multicolumn{2}{c}{Date} & TS & \fermi{} flux &
                                                                   X-ray
                                                                 State\\
	 \cline{1-2}
		(yyyy mm dd) & (MJD) & & (10$^{-7}$ photons cm$^{-2}$s$^{-1}$) & \\
		\hline
          2009-03-05 & 54895 & 10.3 & 4.8\pmi2.0 & HS  \\
          2010-02-02 & 55229 & 10.5 & 6.2\pmi2.3 & HS  \\          
          2010-04-06 & 55292 & 12.2 & 3.1\pmi1.1 & HS \\
          2012-12-31 & 56292 & 9.2 & 5.3\pmi2.2 & SS \\
          2014-02-10 & 56698 & 9.7 & 7.7\pmi3.1 & SS \\
          2014-03-01 & 56717 & 10.5 & 6.3\pmi2.5 & SS \\
          2014-03-06 & 56722 & 9.4 & 6.0\pmi2.5 & HS \\    
          2014-08-08 & 56877 & 10.2 & 6.7\pmi2.5 & SS \\
          2015-05-26 & 57168 & 9.6 & 4.5\pmi1.8 & HS\\
			\hline
	\end{tabular}
        \caption{\label{tab:LC}
          Days (MJD) with a significance $\geq$ 3\,$\sigma$ in the
          1-day bin light curve (0.1--20\,GeV) shown in Figure \ref{fig:LC}.       }
\end{table*}

\section{Discussion}
We established a new statistically significant LAT source above
60\,MeV, spatially coincident with the prototype BH microquasar \cyg,
and the previous marginal detections reported by \citep{Malyshev:2013,
  Sabatini:2010, Sabatini:2013}. 
The use of the more sensitive \pass{} analysis software and 7.5\,yr
of data by \fermi{} allowed us to obtain the first high-significance
detection of a BH binary at high energies, as well as to study a possible
flux variability. The correlation between the HE flux and the hard X-ray
one, together with the hint of flux orbital modulation, strongly
support the identification of the HE source with the microquasar. 
In particular, \cyg{} is detected only when in the HS. During
these periods, the emission is more significant (TS=31)
when the source is around the superior conjunction (0.25$<\phi<$0.75),
while becomes fainter at inferior conjunction (TS=10).
The overall
HS emission is well described by a 2.3 power-law function with a
luminosity of the GeV emission of
$L_{GeV}=5\times10^{33}$\,erg\,s$^{-1}$, few orders of magnitude
smaller than the total power carried by the jets
($10^{36-37}$\,erg\,s$^{-1}$; \citealt{Gallo:2005:nature, Russell:2007}). 

Gamma-ray emission from XRB, and in particular from microquasars, has
been predicted by several authors and associated to either the corona
or the relativistic radio jets. Both leptonic
(e.g. \citealt{Atoyan:1999,georganopoulos:2002}) and hadronic
(\citealt{Romero:2003}) mechanisms have been proposed in the literature to
explain such high-energy radiation (see \citealt{Bosch-Ramon:2009}
for a discussion on different processes). Leptonic models invoke
inverse Compton scattering on seed photons,
where the target photon field depends on the production region, mainly
on the distance from the BH. If particles are accelerated close to the
BH \citep{Kafatos:1981}, the main target photons are the thermal ones
from the accretion disk. When particles are accelerated along the
relativistic jets, the seed photons can be either thermal photons from
the accretion disk or synchrotron soft
photons produced by the same population of electrons
(Synchrotron-Self-Compton, SSC), or the photons from the companion star
(with a black-body peak emission at 2.7$\times$kT$\sim$10\,eV). The
existence of synchrotron emission from the jet is supported by
the hint of strong polarisation in the 0.2--1\,MeV tail
\citep{Laurent:2011, Jourdain:2012, Rodriguez:2015}, the luminosity of
which is $L_{MeV\,tail}\sim7\times10^{35}$\,erg\,s$^{-1}$. 
The inverse Compton scattering on stellar photons would be the 
dominant mechanism of high-energy radiation if the emission is not originated
at the base of the jet. At a distance of few times 10$^{11}$\,cm (see
also \citealt{Romero:2014}), the energy density of the stellar
radiation field ($\omega_{\star}$) becomes dominant with respect to the
other two photon fields. In particular,
$\omega_{\star}=L_{\star}/4\pi(R_{orb}^2+Z^2)c$, where
$L_{\star}$=7$\times$10$^{39}$\,erg\,s$^{-1}$ is the star luminosity
\citep{Orosz:2011},
$R_{orb}$ is the orbital distance assumed to be 3$\times10^{12}$\,cm,
and $Z$ is the distance from the BH along the jet, whereas
$\omega_{synch}=L_{MeV\,tail}/4\pi Z^2 c$ and
$\omega_{accretion}=L_{softXray}/4\pi Z^2 c$, with $L_{softXray}$ the
luminosity in the 1--20\,keV energy range spanning from $10^{36}$ to
$2\times10^{37}$ erg\,s$^{-1}$, depending on the model used to fit the soft
part of the spectrum \citep{DiSalvo:2001}. 
Particles could also
be accelerated outside the binary system in shocks formed when the jets
interact with the surrounding medium, as it is likely to be the case
in the microquasar SS\,433 \citep{Bordas:2015}.  In particular, \cyg{} jets are
thought to inflate a ring-like (5\,pc in diameter) structure, detected at radio
frequencies \citep{Gallo:2005:nature}, and extending 10$^{19}$\,cm away from
the BH. This value is assumed as the maximum extension of the relativistic jets.

If particles are accelerated to relativistic energies close to the BH,
they could create electromagnetic cascades and originate gamma
rays. Those gamma-rays will suffer heavy absorption due to
photon-photon collision. Following the approach of
\citealt{Aharonian:1985b}, we can constrain the
minimum region size for $\sim$GeV photons to escape avoiding pair
production on $\sim$\,1\,keV X-ray photons. Considering a
spherical accretion
geometry, and for a distance of 1.86\,kpc \citep{Orosz:2011} and a
de-absorbed flux at 1\,keV of 1.6$\times10^{-9}$\,erg\,cm$^{-2}$\,s$^{-1}$
\citep{DiSalvo:2001}, the emission region size must be
larger than R$>2\times$10$^9$\,cm. Given that the radius of the
corona, meant as the inner part of the accretion flow, is of
$\sim$20--50\,R$_g\sim$5--10$\times10^7$cm
\citep{Poutanen:1998}, where R$_g$ is the gravitational radius, we can
exclude that the observed GeV emission is produced in the inner
regions of the accretion flow. This absorption also disfavours the
advection-dominated-accretion-flows (ADAF) models, predicting
gamma-ray emission in the HS, when the ADAF flows
are expected to be present \citep{Mahadevan:1997}.  
The GeV emission should be produced
outside the corona, and most likely it is associated with the
jets. This conclusion is also strengthened by the
detection of the system only in the HS, when persistent jets have been
detected at radio frequencies.
Further constraints on the production region can be obtained if the
hint of flux dependence on the orbital phase peaking at around
superior conjunction, as reported in this work, is finally
confirmed. First of all, it sets an upper limit on maximum
distance of the production region of a few times the size of the
system (R$_{orb}$): $Z<10^{13}$\,cm, thus confirming that the GeV
emission cannot come from the region where the jets interact with the
ring-like structure, but from the jets themselves. In addition, such a flux
variability is expected if the production mechanism of this GeV
emission is anisotropic inverse Compton scattering
\citep{jackson:1972,aharonian:1981,zdziarski:2013,khangulyan:2014} on stellar
photons. If no additional sources of variability are assumed, no
variability is expected if either SSC or inverse Compton with the
thermal accretion photons is the dominant mechanism.
Since the energy density of the stellar photons dominates over the
other possible photon fields only at distances $Z>10^{11}$\,cm,
the flux orbital modulation, if confirmed, constrains the GeV emitter
to be located within a $Z$ range 10$^{11}-10^{13}$\,cm.
This region is compatible with the results obtained by hydrodynamic
simulations of stellar winds interacting with 
\cyg-like jets carrying a total power of
$\sim$10$^{36-37}$\,erg\,s$^{-1}$
\citep{Perucho:2008,Yoon:2016,Bosch-Ramon:2016}.

The energy of the parent population of electrons is at least several
tens of GeV, and the inverse Compton scattering occurs mostly in the
Thompson regime. 
A moderate magnetic field of B$\sim10^{-2}$\,G$\times \eta$ (where $\eta$
is the acceleration efficiency: $\tau_{acc}=\eta/qBc$ where
$\tau_{acc}$ is the acceleration timescale, and $\eta>1$) would be
enough to accelerate the inverse Compton
emitting electron population up to a few tens of GeV
\citep{Khangulyan:2008}, enough to produce HE photons via inverse
Compton, in the large stellar photon field. 
Under the assumption that the same population of electrons that
produces the GeV emission by inverse Compton
scattering on stellar photons at $Z=10^{11}-10^{13}$\,cm also emits
synchrotron radiation at lower energies, the maximum 
magnetic field strength in this region is limited by the ratio between
the luminosity of the observed X-ray emission, $L_{X-ray}$,
and the one of the detected high-energy radiation $L_{GeV}$. Otherwise,
the synchrotron X-ray flux would exceed the X-ray observations:
\begin{equation}
\frac{B^2}{8\pi}=\omega_{*}\frac{L_{X-ray}}{L_{GeV}}
\end{equation}
as $L_{X-ray}$ we considered the luminosity between 20 and 100\,keV
of 2.2$\times$10$^{37}$\,erg\,s$^{-1}$ \citep{Cadolle:2006}. At
$Z=10^{12}$\,cm the maximum magnetic field strength is of
$\sim$2\,kG, decreasing down to 700\,G up to $Z=10^{13}$\,cm.

At $\sim 40$\,GeV, the energy spectrum should already be affected by
gamma-ray absorption due to pair creation in the stellar photon
field. The created secondary pairs will mostly radiate inverse Compton
emission around the pair production energy threshold
($\sim10-100$\,GeV) leading, for typical primary
gamma-ray spectra, to the formation of a bump in the SED in that
energy range. The ULs at the highest energies reported in this paper
indicate that the spectrum does not harden above $\sim 10$~GeV. 
If gamma rays are indeed produced at energies $\gtrsim
40$\,GeV, then significant inverse Compton cascading seems unlikely,
which would imply that, either gamma-ray absorption is not attenuated
by electromagnetic cascading in the GeV emitter, or the emission is
produced at the upper end of the inferred emitter Z-range, where this
absorption is expected to be minor  \citep{Bosch-Ramon:2008b}.

In analogy with \cygt, the only other microquasar firmly established
at energies above 100\,MeV, gamma-ray emission of \cyg{} is related to
the existence of relativistic jets. In both cases the GeV emission
is most likely produced by inverse Compton scattering on stellar
photons \citep{Abdo:2009,Dubus:2010}. However, contrarily to \cyg, the
conditions required to detect gamma rays from \cygt{} are that the
source is in the SS and showing significant emission (0.2--0.4\,Jy)
with rapid variations in the radio flux from the radio jets \citep{Corbel:2012}. The
latter is probably related to strong shocks (probably due to discrete
jet ejections) occurring when the source undergoes
state transitions in and out of the ultra-SS. Whereas the nature of
\cygt{} HE emission is transitional, the gamma-ray detection
of \cyg{} seems persistent within the limited statistics,
 whenever the jet is present, i.e. when the source is in the HS, and
 shows a radio flux at 15\,GHz  larger than 10\,mJy\footnote{from 
public AMI-Large Array data. http://www.mrao.cam.ac.uk/$\sim$guy/cx1/} . 
However, the
 daily flux variations weaker than $\sim1.5\times 10^{-6}$
ph\,cm$^{-2}$\,s$^{-1}$ or variations on shorter timescales, as 
reported by the MAGIC collaboration \citep{Albert:2007}, cannot
be excluded (since \cyg{} is not always in the
  field-of-view of the Fermi-LAT on a timescale of an hour). If
confirmed, it could be associated to discrete jet ejections as well,
like in the case of HE emission from \cygt. 

\begin{figure*}[!h]
\centering
\includegraphics[width=0.8\textwidth]{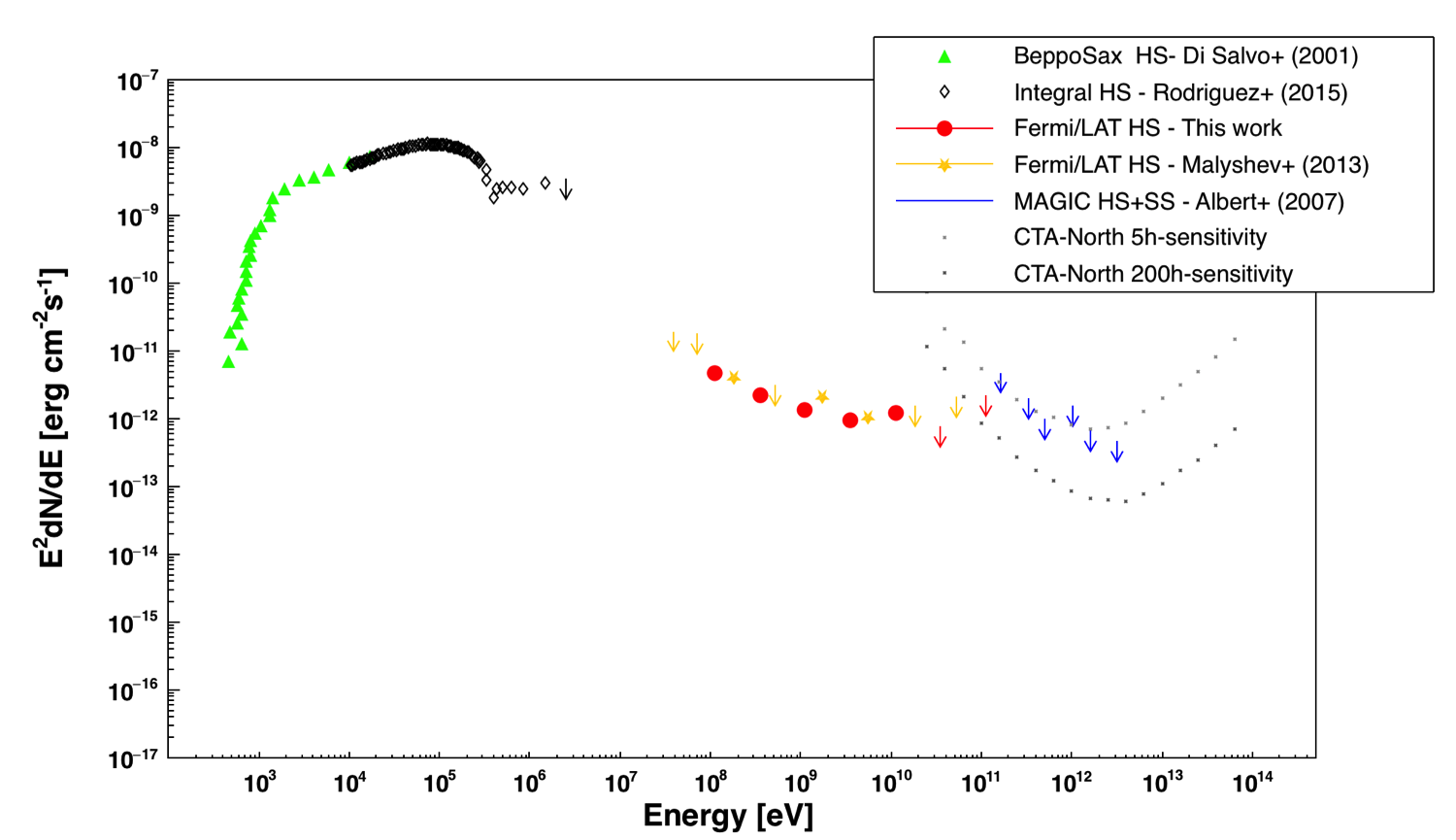}
\caption{Spectral energy distribution of \cyg{} from X-rays up to TeV
  energies when in the HS, except for the MAGIC upper limits which
  were obtained by combining both SS and HS. Soft X-ray (\,keV) data
  are taken from \citet{DiSalvo:2001} (by \emph{BeppoSAX }), hard X-ray
  (10\,keV-2\,MeV) \integral{} from Figure\,3\,in
  \citet{Rodriguez:2015}, and HE (30\,MeV-20\,GeV) results from this work
  and from the previously published ones in \citet{Malyshev:2013}. At
  higher energies the differential UL on the steady emission
  obtained by the MAGIC collaboration \citep{Albert:2007}, under the
  assumption of a 3.2 power law spectrum, are shown. The two gray curves
  are CTA-North differential sensitivities scaled for 5 and 200\,hr of
  observation were taken from the CTA
  webpage
  \emph{https://portal.cta-observatory.org/Pages/CTA-Performance.aspx}.
  No statistical errors are drawn. 
\label{fig:overallSED}}
\end{figure*}

The detection of the spectral break of the HE emission from \cyg{} may
be possible by combining 10\,yr of \fermi{} data and the future
generation of imaging atmospheric Cherenkov telescopes (IACT), the
Cherenkov Telescope Array (CTA). Under the assumption of \pass{} sensitivity,
10\,yr of \fermi{} data will allow to constrain an eventual cutoff
below 100\,GeV, whereas CTA will be more fsensitive above this
energy in 200\,hr of observation. CTA could still detect a different
emission component, possibly of hadronic origin \citep{Pepe:2015}. The sensitivity curve of CTA for
200\,hr of observations with the North array is included in the
broad-band SED shown in Figure\,\ref{fig:overallSED}. Moreover, CTA
will be an optimal instrument to probe the short-term flux variability of \cyg{}
hinted by MAGIC that is showing a flux which one order of magnitude
larger that the ULs obtained for the steady emission. 
Figure\,\ref{fig:overallSED} illustrates the sensitivity curve of the
CTA North array\footnote{Taken from
  https://portal.cta-observatory.org/Pages/CTA-Performance.aspx}
scaled to a few hours of observations, corresponding to the maximum
observation time for a specific source during 1\,d.

\begin{acknowledgements} 
We would like to thank Luigi Tibaldo and Matthew Wood for the fruitful
discussion about the Fermi analysis. We would like also to acknowledge
Tiziana Di Salvo, Guy Pooley and Gabriela Vila for providing us multi-wavelength
historical data, and Marc Rib{\'o} and Pol Bordas for their valuable comments.
R.Z. acknowledges the Alexander von Humboldt Foundation for the
financial support and the Max-Planck Institut fur Kernphysik as
hosting institution. The Spanish funding agency MINECO 
supported the work of O.B. and A.F.B. through the project FPA2012-39502, which
includes ERDF funds, as well as the one of EdOW and D.G.F. and V.B.R.
through the projects AYA2015-71042-P, AYA2013-47447-C3-1-P,
respectively. The Catalan DEC supported EdOW through the grant
SGR2012–1073m and D.G.F. and V.B.R. through the grant SGR2014-86. In
addition, D.G.F. and V.B.R. acknowledge support from MINECO under
grant MDM-2014-0369 of ICCUB (Unidad de Excelencia Mar{\'i}a de
Maezt{\'u}) and BES-2014-069376. EdO and V.B-R. also acknowledges
financial support from MINECO and European Social Funds through a
Ram\'on y Cajal fellowship. Finally, O.B. and A.F.B. are thankful for
the support of the grant SEV-2012-0234 (Centro de Excelencia Severo
Ochoa). This research has been supported by the Marie Curie Career
Integration Grant 321520.

\end{acknowledgements}

\bibliographystyle{aa}
\bibliography{cycX1}


\end{document}